\documentclass[conference]{IEEEtran}
\IEEEoverridecommandlockouts

\usepackage{cite}
\usepackage{amsmath,amssymb,amsfonts}
\usepackage{algorithm}
\usepackage{algorithmic}
\usepackage{graphicx}
\usepackage{textcomp}
\usepackage{xcolor}
\usepackage{acronym}
\usepackage{gensymb}
\usepackage[caption=false]{subfig}
\usepackage{stfloats}
\usepackage{mathrsfs}
\usepackage{multirow}

\usepackage{tikz}
\usetikzlibrary{mindmap, calc, fit, positioning, arrows.meta, shapes.geometric}

\usepackage{amsthm}

\acrodef{ap}[AP]{Access Point}
\acrodef{ai}[AI]{Artificial Intelligence}
\acrodef{fl}[FL]{Federated Learning}
\acrodef{cfl}[CFL]{Clustered Federated Learning}
\acrodef{ml}[ML]{Machine Learning}
\acrodef{nn}[NN]{Neural Network}
\acrodef{sgd}[SGD]{Stochastic Gradient Descent}
\acrodef{svd}[SVD]{Singular Value Decomposition}
\acrodef{cl}[CL]{Centralized Learning}
\acrodef{lstm}[LSTM]{Long Short-Term Memory}
\acrodef{mae}[MAE]{Mean Absolute Error}
\acrodef{cfl-gp}[CFL-GP]{Clustered Federated Learning with Gradient Partitioning}
\acrodef{cfl-2s}[CFL-2S]{Clustered Federated Learning with 2 Stage clustering}
\acrodef{ifca}[IFCA]{Iterative Federated Clustering Algorithm}
\acrodef{ema}[EMA]{Exponential Moving Average}

\def\BibTeX{{\rm B\kern-.05em{\sc i\kern-.025em b}\kern-.08em
    T\kern-.1667em\lower.7ex\hbox{E}\kern-.125emX}}
\begin{document}
\bstctlcite{BSTcontrol} 

\title{An Informativeness-based Clustered Federated Learning Method for Reliable Traffic  Prediction in Managed Wi-Fi Networks
}

\author{\IEEEauthorblockN{Luca Barbieri\textit{$^{+}$}, Gianluca Fontanesi\textit{$^{+}$}, Lorenzo Galati Giordano\textit{$^{+}$}, Alfonso Fernandez Duran$^{\flat}$ and Thorsten Wild\textit{$^{+}$}}
\IEEEauthorblockA{$^{+}$\emph{Radio Systems Research, Nokia Bell Labs, Stuttgart, Germany}}
\IEEEauthorblockA{$^{\flat}$\emph{Nokia Spain, Madrid}}
\thanks{This work is partially supported by UNITY-6G project, co-funded from European Union’s Horizon Europe Smart Networks and Services Joint Undertaking (SNS JU) research and innovation programme under the Grant Agreement No 101192650 and from the Swiss State Secretariat for Education, Research and Innovation (SERI). This work is also partially supported by CDTI project IDI-20250211 MINERGY.}}

\maketitle

\begin{abstract} 
Centrally-managed Wi-Fi solutions are increasingly leveraging Distributed Artificial Intelligence (AI) to predict key operational statistics of Access Points (APs) and proactively optimize network performance. 
In this context, Clustered Federated Learning (CFL) represents a fitting methodology, enabling the generation of multiple AI models that account for diverse statistical properties of the APs data distribution. 
However, identifying informative clusters for grouping APs models remains a significant challenge. 
In this paper, we address this problem by proposing a novel CFL tool integrating a two step clustering procedure.
Initially, multiple clustering solutions are generated and filtered based on a minimum set of desired clustering criteria.
Subsequently, if no solutions meet sufficient quality metrics, a global model is produced by aggregating all AP models. 
Otherwise, the final clustering solution is selected as the one that maximizes the informativeness (quantified via differential entropy) for the smallest cluster.
Our results, focusing on a Wi-Fi traffic prediction problem, demonstrate that the developed CFL tool achieves the best predictive performance among all evaluated distributed strategies and the lowest communication and energy footprint among the clustered ones, exceeding the cost of single-model FL only in the regimes where it markedly improves accuracy.   
\end{abstract}


\section{Introduction}
\label{sec:intro}

Wi-Fi is the key radio technology for delivering indoor connectivity across residential, enterprise, and public environments, favored for its ease of deployment and ubiquity~\cite{wifi-survey,survey-lorenzo}.
Beyond traditional applications, Wi-Fi has evolved to support a wide variety of emerging and challenging use cases, including immersive communications~\cite{xr_wifi}, digital twins for manufacturing~\cite{digital_twins_wifi} and healthcare~\cite{heathcare_wifi}.
Meeting these demands has driven recent standard amendments (e.g., Wi-Fi~8~\cite{wifi8_primer}) while inevitably raising network complexity and motivating novel orchestration paradigms.
A prominent paradigm gaining traction is managed Wi-Fi~\cite{managed_wifi}, where operators leverage centralized controllers to optimize \acp{ap} radio parameters and ensure seamless connectivity through the continuous monitoring of key operational statistics collected from the network.

The resulting data complexity calls for advanced analytics, and distributed \ac{ai}-driven methods are increasingly adopted to support Wi-Fi
management~\cite{wifi_meets_ml}. Among these, \ac{fl}~\cite{fl} is a particularly fitting paradigm, as it learns a shared global model collaboratively across the
\acp{ap} without centralizing raw data. A single global model, however, is often
suboptimal in managed Wi-Fi, where heterogeneous data distributions across the \acp{ap} hinder convergence to a solution that performs well for all. \ac{cfl}~\cite{cfl,cfl_survey2}
overcomes this limitation by partitioning the \acp{ap} into groups of similar clients and learning a dedicated per-cluster model, making it a natural fit for managed Wi-Fi systems. 
The central challenge then becomes identifying informative \ac{ap} clusters from the federated updates alone

Clustering has been investigated in managed Wi-Fi, albeit only at the cloud controller and over centrally collected data~\cite{gianluca_wifi}, thereby reintroducing the data centralization that \ac{fl} is designed to avoid. 
On the federated side, prior \ac{cfl} efforts in Wi-Fi have targeted fingerprinting, sensing, and intrusion detection~\cite{cfl_on_wifi,cfl_on_wifi2,cfl_wifi3}, whereas traffic prediction has relied on vanilla \ac{fl} with a single global model~\cite{fl_traffic,fl_traffic2} that overlooks per-\ac{ap} heterogeneity. 
To the best of our knowledge, this is the first work to integrate clustering within \ac{fl} for traffic prediction in managed Wi-Fi.
 
This paper advocates leveraging \ac{cfl} in managed Wi-Fi to enhance predictive performance. 
To this end, we propose a novel \ac{cfl} methodology built on a two-stage clustering algorithm, referred to as \ac{cfl-2s}, which produces more informative clusters than state-of-the-art approaches. 
In the first stage, the method identifies a set of candidate partitions that satisfy a minimum set of clustering criteria. 
In the second, it selects the partition that maximizes the informativeness of its smallest cluster, quantified by the differential entropy of the per-\ac{ap} gradients. 
This focus is motivated by the fact that the smallest cluster contains the fewest \acp{ap} and hence the least training data, making it the
most prone to poor generalization. Crucially, the method operates solely on the gradient updates already exchanged during \ac{fl}, rendering it agnostic to the learning task, model architecture, and data modality. 
Thus, it applies broadly to heterogeneous federated settings. 
We validate \ac{cfl-2s} on a real-world managed Wi-Fi dataset for traffic prediction, where it attains the best predictive performance among all distributed strategies, namely up to $51\%$ and $21\%$ improvement over vanilla \ac{fl} and state-of-the-art \ac{cfl} baselines, respectively.
Besides, it incurs the lowest communication overhead and
energy among the \ac{cfl} methods.

The rest of the paper is organized as follows. Sec.~\ref{sec:sys_model} details the federated system model for traffic prediction, while Sec.~\ref{sec:cfl} presents the proposed \ac{cfl} solution, whose results are analyzed in Sec.~\ref{sec:results}.
Finally, Sec.~\ref{sec:conclusions} draws some conclusions.

\section{System model}
\label{sec:sys_model}

This section introduces the system model underpinning the proposed \ac{cfl} approach.
Sec.~\ref{subsec:load_pred_problem} formulates the traffic prediction problem both for centralized and federated settings, while Sec.~\ref{subsec:cluster_based_solution} details the \ac{fl} protocol.

\begin{figure}
    \centering
    \includegraphics[width=\linewidth]{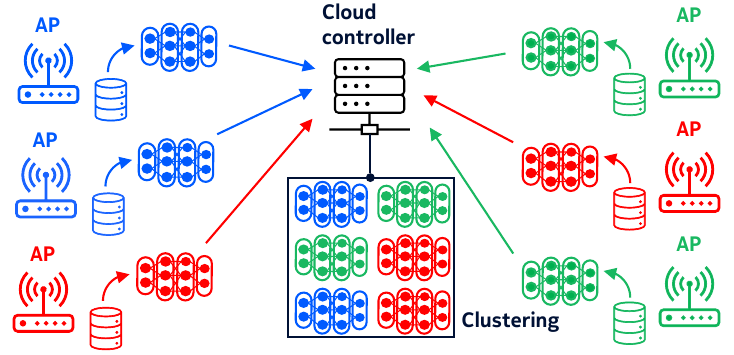}
    \vspace{-8mm}
    \caption{\ac{cfl} system: \acp{ap} train a local copy of a shared \ac{ml} model which is shipped to the cloud controller for clustering similar models together.}
    \label{fig:cfl_system}
\end{figure}

\subsection{Traffic prediction problem}
\label{subsec:load_pred_problem}

We consider a managed Wi-Fi system comprising a set $\mathcal{A} = \{1, \ldots, N_{\text{AP}} \}$ of $N_{\text{AP}}$ \acp{ap} managed by a cloud controller, as exemplified in Fig. \ref{fig:cfl_system}.  
Each \ac{ap} $a \in \mathcal{A}$ collects time-series data from its radio interfaces describing the traffic evolution over time $t \in \{0, \ldots, \text{T}\}$, where $\text{T}$ denotes the current time.
Time is sampled at discrete events, with the interval between successive time instants ruled by the system time granularity $\text{T}_{\text{s}}$ e.g., hourly or on a per-minute basis). 
This data for the $a$-th \ac{ap} can be compactly described as $\mathbf{X}_{0:\text{T}}^{(a)} = \{\mathbf{x}_{0}^{(a)}, \ldots, \mathbf{x}_{\text{T}}^{(a)} \}$, where $\mathbf{x}_{t}^{(a)}$ is a vector at time $t$ comprising $M$ features, which include traffic metrics and additional context information~\cite{bbf} useful for prediction (e.g., number of active connections, current time).
For solving the traffic prediction problem, each \ac{ap} $a \in \mathcal{A}$ builds a local dataset via a sliding window of length $\text{L} + \text{H}$, where $\text{L}$ is the input context length and $\text{H}$ the prediction horizon as
\begin{equation}
    \mathcal{D}^{(a)} = \left\{ \left(\mathbf{X}_{n:n+\text{L}}^{(a)},\, \mathbf{X}_{n+\text{L}+1:n+\text{L}+\text{H}}^{(a)}\right) \right\}_{n=1}^{N_{\text{E}}^{(a)}} \, ,
    \label{eq:local_dataset}
\end{equation}
where $N_{\text{E}}$ is the number of samples. 

For centralized solutions, the controller collects all local datasets, forming $\mathcal{D} = \bigcup_{a \in \mathcal{A}} \mathcal{D}^{(a)}$ with $N_{\text{E}} = \sum_{a \in \mathcal{A}} N_{\text{E}}^{(a)}$ total samples and learns a single shared mapping $f_{\text{C}}(\cdot)$ as
\begin{equation}
\widehat{\mathbf{X}}_{t:t+\text{H}}^{(a)} = f_{\text{C}}\!\left(\mathbf{X}_{t-\text{L}:t-1}^{(a)}\right) \quad \forall\, a \in \mathcal{A}, \, \forall\, t \in \{0, \ldots, \text{T}\} \, .
\end{equation}
This mapping is parameterized by the weights $\mathbf{W}$ of a \ac{nn} learned by minimizing 
\begin{multline}
    \min\limits_{\mathbf{W}} \mathscr{L}(\mathbf{W}) = \min\limits_{\mathbf{W}} \frac{1}{N_{\text{E}}} \sum_{a \in \mathcal{A}} \sum_{n = 1}^{N_{\text{E}}^{(a)}} \\
    \ell\!\left(\mathbf{X}_{n+\text{L}+1:n+\text{L}+\text{H}}^{(a)}, \widehat{\mathbf{X}}_{n+\text{L}+1:n+\text{L}+\text{H}}^{(a)}; \mathbf{W}\right) \, ,
    \label{eq:global_loss}
\end{multline}
where $\ell(\cdot)$ denotes the loss function (e.g., L1 loss) evaluated over each sample in $\mathcal{D}$. 

Nevertheless, centralizing raw data constrains the attainable temporal
granularity, since fine-grained sampling entails frequent, high-volume transfers
between the \acp{ap} and the controller. \ac{fl} mitigates this by exchanging
model parameters rather than raw data, thereby relaxing the granularity
constraint while offloading part of the computational burden from the cloud.

\subsection{Federated-based solution}
\label{subsec:cluster_based_solution}

The \ac{fl} goal is amenable to the optimization in \eqref{eq:global_loss}, where now the objective is to minimize the following function
\begin{equation}
    \min\limits_{\mathbf{W}} \mathcal{L}(\mathbf{W}) = \min\limits_{\mathbf{W}} \sum_{a \in \mathcal{A}} \rho_a \mathcal{L}_{a}(\mathbf{W}) \, , 
\end{equation}
with $\rho_a =  N_{\text{E}}^{(a)}/N_{\text{E}}$ and
\begin{equation}
    \mathcal{L}_{a}(\mathbf{W}) = \! \dfrac{1}{N_{\text{E}}^{(a)}} \! \sum_{n=1}^{N_{\text{E}}^{(a)}} \!\ell\!\left(\mathbf{X}_{n+\text{L}+1:n+\text{L}+\text{H}}^{(a)}, \widehat{\mathbf{X}}_{n+\text{L}+1:n+\text{L}+\text{H}}^{(a)}; \mathbf{W}\right),  \label{eq:fl_loss} 
\end{equation}
where $\ell(\cdot)$ is the same loss function as in \eqref{eq:global_loss}, now evaluated over the local dataset $\mathcal{D}^{(a)}$. 
Solving \eqref{eq:fl_loss} in a federated manner is done by alternating local optimization steps at the \acp{ap} with \ac{ml} model aggregation at the cloud controller so as to learn a resilient global model.
More formally, at each iteration $i$, each \ac{ap} $a \in \mathcal{A}$ updates its local copy of the \ac{ml} model $\mathbf{W}_{i}^{(a)}$ via stochastic gradient steps as follows
\begin{equation}
    \mathbf{W}_{i,a}^{'} = \mathbf{W}_{i,a} - \eta \nabla \mathcal{L}_{a}(\mathbf{W}_{i,a}) \, , \label{eq:local_sgd}
\end{equation}
where $\eta$ is the learning rate. 
\acp{ap} then evaluate the surrogate gradient 
\begin{equation}
    \boldsymbol{\Delta}_{i,a} = (\mathbf{W}_{i,a} - \mathbf{W}_{i,a}^{'})/\eta \, ,
\end{equation}
and share it with the cloud controller for aggregation. 
The global model is then updated as
\begin{equation}
    \mathbf{W}_{i+1} = \mathbf{W}_{i} - \eta \sum_{a \in \mathcal{A}_{0}} \sigma_a \boldsymbol{\Delta}_{i,a} \, ,  
\end{equation}
where $\mathcal{A}_{0}$ denotes the set of \acp{ap} that shared their updates to the cloud controller, while 
\begin{equation}
    \sigma_a = \dfrac{N_{\text{E}}^{(a)}}{\sum_{a' \in \mathcal{A}_{0}} N_{\text{E}}^{(a')}} \, , 
\end{equation}
is a weighting factor.
\acp{ap} then substitute their local model with the received global one and a new iteration begins, until $I_{\max}$ rounds are reached.

While \ac{fl} is attractive for managed Wi-Fi, a single global model may underperform in the presence of heterogeneous \ac{ap} traffic distributions, motivating the \ac{cfl} strategy detailed next.

\section{Clustered federated learning method}
\label{sec:cfl}

This section details the novel two-stage \ac{cfl} method that generalizes the \ac{fl} optimization of Sec.~\ref{subsec:cluster_based_solution} to a per-cluster objective (Sec.~\ref{subsec:cfl_objective}).
Stage~1 (Sec.~\ref{subsec:stage1}) generates and filters candidate \ac{ap} partitions via \ac{svd} and $k$-means; 
Stage~2 (Sec.~\ref{subsec:stage2}) selects the most informative one, until the optimal cluster count $K^{\dagger}$ is found---after which re-clustering simply reruns \ac{svd} and $k$-means with $K^{\dagger}$ fixed.

\subsection{From federated to clustered federated learning}
\label{subsec:cfl_objective}

Let $\mathcal{P} = \{\mathcal{C}_1, \ldots, \mathcal{C}_K\}$ denote a partition of the \acp{ap} set $\mathcal{A}$ into $K$ disjoint clusters, i.e., $\bigcup_{k=1}^{K} \mathcal{C}_k = \mathcal{A}$ and $\mathcal{C}_k \cap \mathcal{C}_{k'} = \emptyset$ for $k \neq k'$. 
Differently from the vanilla \ac{fl} formulation in \eqref{eq:fl_loss}, in \ac{cfl} a distinct \ac{ml} model $\mathbf{W}^{(k)}$ is learned for each cluster $\mathcal{C}_k$ by solving
\begin{equation}
    \min\limits_{\mathbf{W}^{(k)}} \mathcal{L}^{(k)}(\mathbf{W}^{(k)}) = \min\limits_{\mathbf{W}^{(k)}} \sum_{a \in \mathcal{C}_k} \rho_a^{(k)} \mathcal{L}_{a}(\mathbf{W}^{(k)}) \, , 
    \label{eq:cfl_loss}
\end{equation}
with cluster-wise data weighting
\begin{equation}
    \rho_a^{(k)} = \dfrac{N_{\text{E}}^{(a)}}{\sum_{a' \in \mathcal{C}_k} N_{\text{E}}^{(a')}} \, .
    \label{eq:cfl_rho}
\end{equation}
The local update rules of Sec.~\ref{subsec:cluster_based_solution} carry over unchanged at the \acp{ap}, while the cloud controller now performs $K$ separate aggregations, one per cluster, namely
\begin{equation}
    \mathbf{W}_{i+1}^{(k)} = \mathbf{W}_{i}^{(k)} - \eta \sum_{a \in \mathcal{A}_0 \cap \mathcal{C}_k} w_a^{(k)} \boldsymbol{\Delta}_{i,a} \, ,
    \label{eq:cfl_aggregation}
\end{equation}
with $w_a^{(k)} = N_{\text{E}}^{(a)} / \sum_{a' \in \mathcal{A}_0 \cap \mathcal{C}_k} N_{\text{E}}^{(a')}$.
Each \ac{ap} $a \in \mathcal{C}_k$ then synchronizes its local copy with $\mathbf{W}_{i+1}^{(k)}$ before the next round.
The central question becomes how to compute a partition $\mathcal{P}$ that yields informative and well-separated groups of \acp{ap}.
We address this through a two-stage search procedure that starts from a single global model ($\mathcal{P}=\{\mathcal{A}\}$, $K^{\dagger}=\varnothing$), which serves as the effective warm-up until Stage~2 first identifies a valid partition and sets $K^{\dagger}$. 
From that point onwards, periodic re-clustering at every $K_{\text{rc}}$ rounds simply refreshes the \ac{ap} assignments via \ac{svd} and $k$-means with $K^{\dagger}$ fixed.

\subsection{Stage 1: candidate partitions via SVD and \texorpdfstring{$k$}{k}-means}
\label{subsec:stage1}

To reduce the sensitivity to noise in individual-round gradients, the controller maintains a per-\ac{ap} gradient feature updated as an \ac{ema} after every \ac{fl} round $i$ as
\begin{equation}
\mathbf{g}_{i,a} = (1-\beta)\, \mathbf{g}_{i-1,a} + \beta\, \Delta_{i,a} \, , \quad \mathbf{g}_{0,a} = \mathbf{0} \, ,
    \label{eq:grad_ema}
\end{equation}
with smoothing factor $\beta \in (0,1]$.
At a re-clustering round $i_r$, the cloud controller stacks these accumulated gradient features as
\begin{equation}
\mathbf{D}_{i_r} = \big[\, \mathbf{g}_{i_r,1} \;\; \mathbf{g}_{i_r,2}, \ldots, \mathbf{g}_{i_r,N_{\text{AP}}} \,\big] \, .
    \label{eq:gradient_matrix}
\end{equation}
To mitigate the high dimensionality of $\mathbf{D}_{i_r}$, the controller computes its thin \ac{svd}
\begin{equation}
    \mathbf{D}_{i_r} = \mathbf{U} \boldsymbol{\Sigma} \mathbf{V}^{\top} \, , 
    \label{eq:svd}
\end{equation}
where the columns of $\mathbf{U}$ are the left singular vectors, $\mathbf{V}$ the right singular vectors, and $\boldsymbol{\Sigma}$ the diagonal matrix of singular values $\sigma_1 \geq \cdots \geq \sigma_R \geq 0$.
The truncation rank is selected as $r = \mathrm{clip}\!\left(1 + \arg\max_{k \in \{1,\ldots,R-1\}}(\sigma_k - \sigma_{k+1}),\, 3,\, N_{\text{AP}}\right)$, i.e., the position of the largest spectral gap clipped to $[3, N_{\text{AP}}]$ to avoid degenerate single-direction embeddings and ranks exceeding the number of \acp{ap}. The low-dimensional \ac{ap} embedding matrix is then obtained as 
\begin{equation} 
    \widetilde{\mathbf{D}}_{i_r} = \mathbf{U}_{r}^{\top} \mathbf{D}_{i_r} = \boldsymbol{\Sigma}_{r} \mathbf{V}_{r}^{\top} \, , 
    \label{eq:embedding}
\end{equation}
where the $a$-th column is the $a$-th \ac{ap} low-dimensional embedding.

A pool of candidate partitions is then generated by sweeping the number of clusters from $2$ up to $K_{\max}$, set as $K_{\max}=r$ 
since the $r$ dominant directions capture the heterogeneity modes in the gradient profiles; partitions with $K > r$ rely on noise-dominated directions and should be discarded. The lower clip at $r=3$ ensures the candidate set always spans at least two distinct partition sizes.
This yields the candidate set
\begin{equation}
    \mathfrak{P} = \big\{ \mathcal{P}_K : K \in \{2, \ldots, K_{\max}\} \big\} \, , 
    \label{eq:candidate_set}
\end{equation}
where $\mathcal{P}_K$ is the partition produced by running $k$-means on the columns of $\widetilde{\mathbf{D}}_{i_r}$ with $K$ centroids. 
Each candidate partition is then assessed against a set of quality criteria $\mathcal{Q}$, which may encode geometric measures (e.g., cluster separation or cohesion) as well as structural constraints (e.g., minimum cluster size); see Sec.~\ref{sec:results} for the specific instantiation used in this work.
Only partitions satisfying all criteria in $\mathcal{Q}$ are retained, yielding the filtered set
\begin{equation}
    \mathfrak{P}^{\star} = \big\{ \mathcal{P} \in \mathfrak{P} : \mathcal{P} \text{ satisfies } \mathcal{Q} \big\} \, .
    \label{eq:filtered_set}
\end{equation}
If $\mathfrak{P}^{\star} = \emptyset$, no candidate satisfies the desired clustering quality and the procedure falls back to a single global model produced by the standard \ac{fl} aggregation of Sec.~\ref{subsec:cluster_based_solution}.
The re-clustering attempt is then deferred to the next window of $K_{\text{rc}}$ rounds.

\subsection{Stage 2: informativeness-driven partition selection}
\label{subsec:stage2}

Whenever $\mathfrak{P}^{\star}$ is non-empty, Stage~2 selects the partition that yields the most informative smallest cluster.
This focus is motivated by the observation that small clusters gather the fewest \acp{ap} and therefore the least training data, which directly widens their generalization gap.
Partitions whose smallest cluster carries highly informative gradient profiles are therefore preferred, as they reduce the risk of producing under-trained clustered models.

\begin{algorithm}[t]
\caption{Proposed two-stage \ac{cfl} procedure}
\label{alg:cfl}
\begin{algorithmic}[1]
\REQUIRE \acp{ap} set $\mathcal{A}$, period $K_{\text{rc}}$, quality criteria $\mathcal{Q}$, \ac{ema} factor $\beta$
\STATE Initialize $\mathcal{P} \leftarrow \{\mathcal{A}\}$, $\mathbf{W}^{(1)}$, $g_{0,a} \leftarrow \mathbf{0}$ $\forall\, a \in \mathcal{A}$, $K^{\dagger} \leftarrow \varnothing$
\FOR{$i = 1, 2, \ldots, I_{\max}$}
    \FORALL{$a \in \mathcal{A}$}
        \STATE Update $\mathbf{W}_{i,a}$ as in \eqref{eq:local_sgd} and send $\Delta_{i,a}$
        \STATE Set $g_{i,a} \leftarrow (1-\beta)\,g_{i-1,a} + \beta\,\Delta_{i,a}$
    \ENDFOR
    \STATE Cloud aggregates per cluster as in \eqref{eq:cfl_aggregation}  
    \IF{$i \mod K_{\text{rc}} = 0$}
        \STATE Compute SVD embeddings $\widetilde{\mathbf{D}}_{i}$ as in \eqref{eq:gradient_matrix}-\eqref{eq:embedding} and set $K_{\max} \leftarrow r$
        \IF{$K^{\dagger} = \varnothing$}
            \STATE Sweep $k$-means over $K\in\{2,\ldots,K_{\max}\}$ on $\widetilde{\mathbf{D}}_{i}$
            \STATE Filter $\mathfrak{P}^{\star}$ via \eqref{eq:filtered_set}
            \IF{$\mathfrak{P}^{\star} = \emptyset$}
                \STATE Revert to global model: $\mathcal{P} \leftarrow \{\mathcal{A}\}$
            \ELSE
                \STATE For each $\mathcal{P}\in\mathfrak{P}^{\star}$, compute $\mathcal{H}(\mathcal{P})$ via \eqref{eq:per_ap_entropy}--\eqref{eq:cluster_entropy}
                \STATE Set $\mathcal{P} \leftarrow \mathcal{P}^{\dagger}$ as in \eqref{eq:partition_selection} and $K^{\dagger} \leftarrow |\mathcal{P}^{\dagger}|$
            \ENDIF
        \ELSE
            \STATE Run $k$-means with $K{=}K^{\dagger}$ on $\widetilde{\mathbf{D}}_{i}$ to refresh $\mathcal{P}$
        \ENDIF
    \ENDIF
\ENDFOR
\end{algorithmic}
\end{algorithm}

For a partition $\mathcal{P} \in \mathfrak{P}^{\star}$, let $\mathcal{C}_{\star}(\mathcal{P}) = \arg\min_{k \in \{1,\ldots,K\}} |\mathcal{C}_k|$ denote its smallest cluster (ties broken arbitrarily).
For each \ac{ap} $a \in \mathcal{C}_{\star}(\mathcal{P})$, the Gaussian differential entropy of the entries of $\mathbf{g}_{i_r}^{(a)}$ is used as a hyperparameter-free informativeness score
\begin{equation}
    H^{(a)} = \frac{1}{2}\ln\!\left(2\pi e\,\hat{\sigma}^{(a)2}\right) , \quad
    \hat{\sigma}^{(a)2} = \frac{1}{P}\sum_{p=1}^{P}\left(g_{i_r,p}^{(a)} - \bar{g}_{i_r}^{(a)}\right)^2 \, ,
    \label{eq:per_ap_entropy}
\end{equation}
where $\bar{g}_{i_r}^{(a)}$ is the mean gradient entry and $\hat{\sigma}^{(a)2}$ its empirical variance.
Since the Gaussian is the maximum-entropy distribution over the reals for a fixed variance~\cite{cover_thomas}, this score is a conservative informativeness measure requiring no assumption beyond the empirical second moment of the gradient entries.
The overall informativeness of partition $\mathcal{P}$ is then defined as the cumulative entropy of its smallest cluster,
\begin{equation}
    \mathcal{H}(\mathcal{P}) = \sum_{a \in \mathcal{C}_{\star}(\mathcal{P})} H^{(a)} \, .
    \label{eq:cluster_entropy}
\end{equation}
The final partition retained for clustered aggregation is the one maximizing this score,
\begin{equation}
    \mathcal{P}^{\dagger} = \arg\max\limits_{\mathcal{P} \in \mathfrak{P}^{\star}} \mathcal{H}(\mathcal{P}) \, ,
    \label{eq:partition_selection}
\end{equation}
with ties resolved in favor of the partition exhibiting the largest minimum-size cluster, so as to additionally promote balanced cluster cardinalities.
The selected partition $\mathcal{P}^{\dagger}$, with $K^{\dagger} = |\mathcal{P}^{\dagger}|$ clusters, marks the end of the \emph{search phase}.
Subsequently, the controller enters a \emph{maintenance phase}: at each re-clustering round, it reapplies \ac{svd} and $k$-means with $K = K^{\dagger}$ fixed to refresh the \ac{ap} assignments, without re-executing Stage~2, thereby avoiding the cost and the cluster-count oscillation that repeatedly re-running the clustering operations search across rounds would otherwise introduce.

\section{Numerical results}
\label{sec:results}

This section evaluates the performance of the proposed \ac{cfl-2s} against several baselines.
Sec.~\ref{subsec:dataset} introduces the dataset and the main simulation parameters.
Sec.~\ref{subsec:traffic_results} and Sec.~\ref{subsec:energy_comm_results} discuss prediction performance and energy and communication overhead, respectively. 

\subsection{Dataset and simulation parameters}
\label{subsec:dataset}  

Our experiments build on the open-source association records of~\cite{flags}, which
log 7{,}404 Wi-Fi \acp{ap} deployed across a 3~km$^2$ campus over 49 days in 2019.
Following~\cite{fl_traffic}, we post-process these records to derive the per-AP uplink and
downlink traffic time series that constitute the target of our prediction task.
The goal is to forecast the \ac{ap} traffic over $H\in\{1,6\}$ future time steps (corresponding to 10-min and 1-hour ahead, respectively) given the preceding $\text{L}=60$ observations (i.e., the preceding 10~hours), at a temporal resolution of $\text{T}_{\text{s}}=10$~minutes. For each evaluated network size $N_{\text{AP}} \in \{10, 50, 100\}$, \acp{ap} are drawn uniformly at random from the full dataset, with the same random seeds used across all learning strategies to ensure a fair comparison.
 
We compare \ac{cfl-2s} against: (i)~vanilla \ac{fl} (Sec.~\ref{sec:sys_model}); (ii)~\ac{ifca}~\cite{ifca}, which alternates cluster assignment and per-cluster updates with a fixed $K$ set equal to the $K^{\dagger}$ identified by \ac{cfl-2s} for each run (an oracle cluster count not available in practice); and (iii)~\ac{cfl-gp}~\cite{cfl_sota}, a gradient-based \ac{cfl} state-of-the-art baseline.
All strategies employ an \ac{lstm} model comprising one hidden layer of 50 units followed by a fully connected output layer of $H$ neurons, with ${\sim}11$k parameters.
Training is carried out for up to $I_{\max}=300$ rounds using the Adam optimizer ($\eta=0.001$, mini-batch size 32), with early stopping applied if performances do not improve over 20 consecutive communication rounds.
An 80\%/20\% train-test split is adopted, applied chronologically per AP: the first 80\% of each \ac{ap}'s time windows form the training set and the remaining 20\% the test set, preserving temporal ordering. 
All results are averaged over 5 independent random seeds to account for variability.
Regarding the clustering-specific hyperparameters, both \ac{cfl} approaches share a smoothing factor $\beta=0.9$ and a re-clustering period of $K_{\text{rc}}=2$ communication rounds.
For the quality criteria $\mathcal{Q}$ (Sec.~\ref{subsec:stage1}), we adopt a single minimum silhouette threshold $s_{\min}=0.7$. The silhouette score~\cite{rousseeuw1987} measures, for each \ac{ap}, the ratio of inter-cluster separation to intra-cluster cohesion, with the per-partition score $s(\mathcal{P}) \in [-1,+1]$ (where $+1$ indicates perfect separation) being the mean over all \acp{ap}. This criterion is chosen as it reuses the embedding already computed in Stage~1 and captures both cluster compactness and separation with a single, interpretable scalar. The value $s_{\min}=0.7$ lies in the ``strong structure'' region of the silhouette scale~\cite{rousseeuw1987}), ensuring that only partitions with well-separated, cohesive clusters are retained. A systematic study of alternative quality metrics and thresholds is beyond our scope and left to future work.

Four metrics are considered: (i)~\ac{mae} for prediction quality; (ii)~convergence rounds: the round achieving the best \ac{mae} before early stopping or $I_{\max}$ is reached; (iii)~communication overhead (MB), accounting for the total data exchanged between \acp{ap} and the cloud controller up to the convergence round, including gradient uploads and model downloads at each round; and (iv)~computational energy consumed up to the convergence round, measured via \textit{Eco2ai}~\cite{eco2ai}\footnote{Simulations were executed on a PC equipped with an Intel(R) Xeon(R) W3-2435 CPU and 256 GB of RAM. Energy measured on a CPU; a GPU would change absolute consumption but is expected to preserve the relative ranking across strategies, as all run the same model and number of rounds.} (this reflects algorithmic cost and does not include wireless transmission energy at the \acp{ap}). 

\subsection{Traffic prediction quality assessment}
\label{subsec:traffic_results} 

\begin{figure}[!t]
    \includegraphics[width=\linewidth]{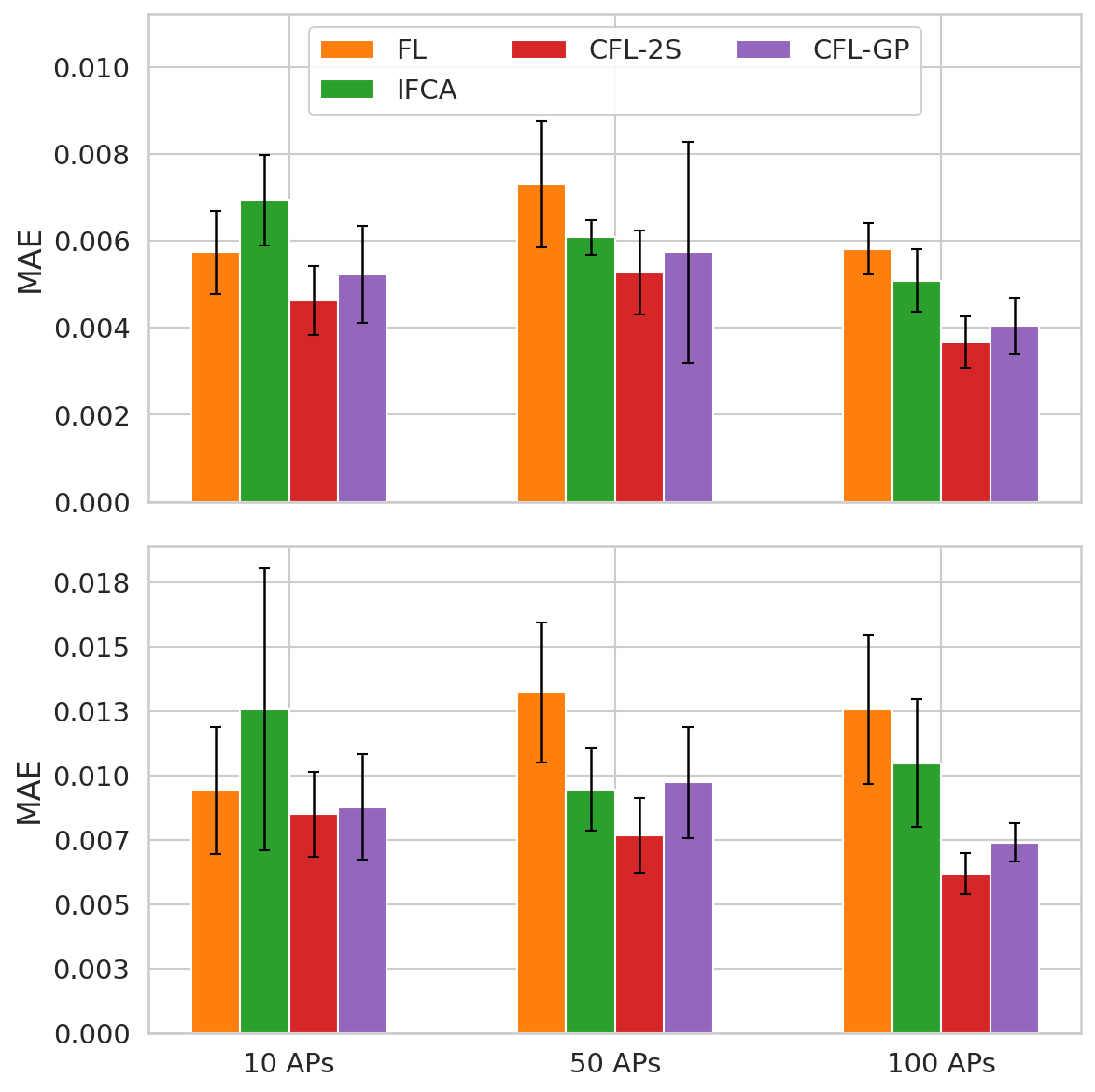}
    \vspace{-7mm}
    \caption{Prediction performance in terms of \ac{mae} for each learning strategy; $\text{H}=1$ top and $\text{H}=6$ bottom; error bars represent $\pm 1$ std over 5 seeds.}
    \label{fig:mae_results}
\end{figure}  

Fig.~\ref{fig:mae_results} reports the \ac{mae} achieved by each strategy across $N_{\text{AP}} \in \{10, 50, 100\}$ and both prediction horizons.
Analyzing the results, vanilla \ac{fl} underperforms due to its single global model, while \ac{ifca}, despite relying on the same oracle cluster count $K^{\dagger}$ as \ac{cfl-2s}, achieves comparably poor \ac{mae}. This might suggest that having access only to the optimal cluster count alone is insufficient to improve performances, and a better cluster selection, as the one employed by our proposal, is what brings the most significant gains.
\ac{cfl-gp} provides moderate \ac{mae} reductions over \ac{fl} and \ac{ifca} through per-cluster modeling, though its gains are less consistent across network sizes and prediction horizons.
On the other hand, the proposed \ac{cfl-2s} method achieves the best \ac{mae} among all distributed strategies across every evaluated configuration, with the largest improvements at the longer horizon $\text{H}=6$ and larger networks, where it reduces the \ac{mae} by up to $51\%$ over vanilla \ac{fl} and $21\%$ over \ac{cfl-gp}. At the short horizon the gains shrink and the gap to \ac{cfl-gp} narrows, falling within one standard deviation at the largest network, 
Still, our proposal attains the lowest mean \ac{mae} throughout.

\subsection{Convergence, communication and energy analysis}
\label{subsec:energy_comm_results}

\begin{figure}[!t]
    \includegraphics[width=\linewidth]{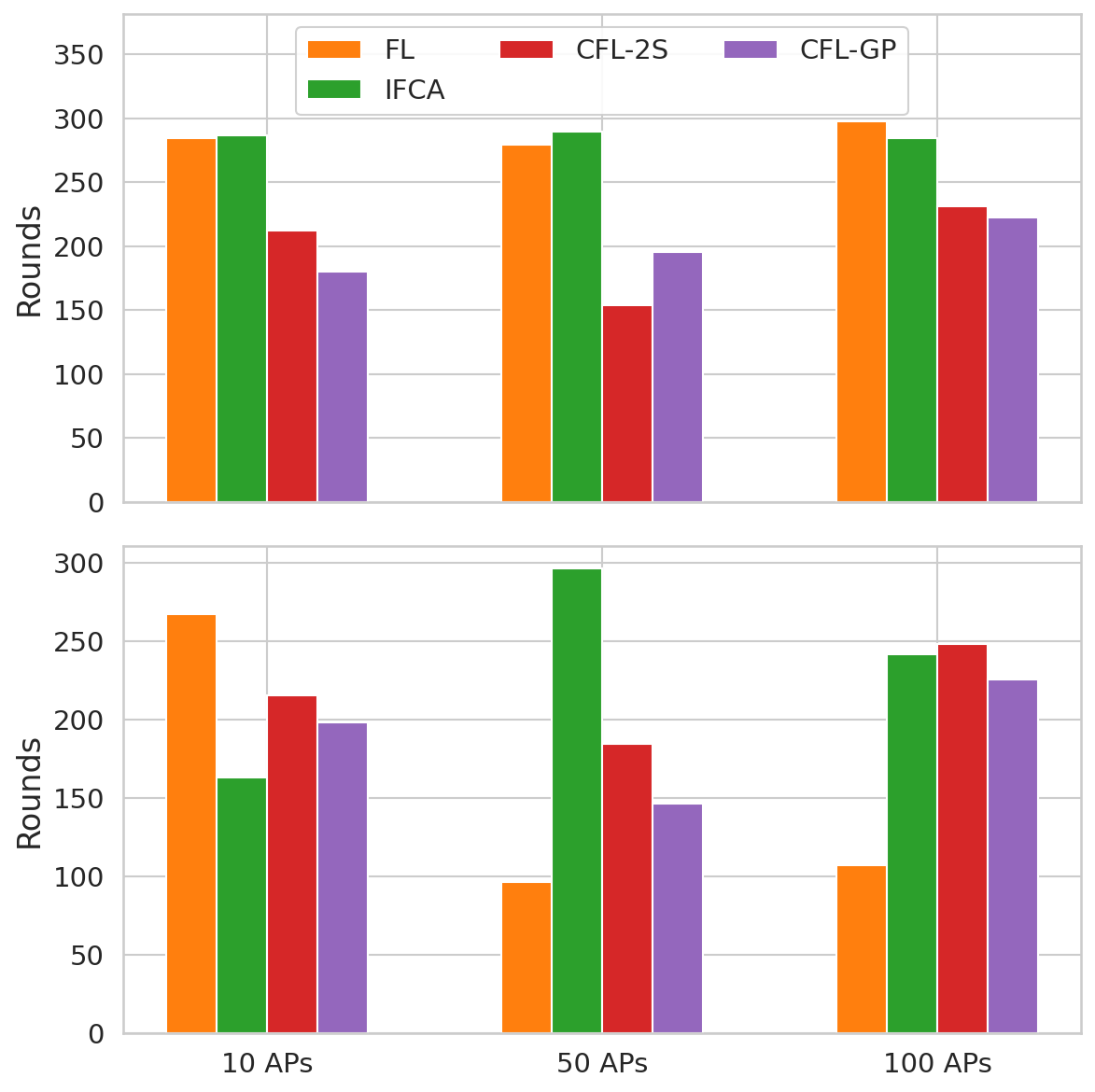}
    \vspace{-7mm}
    \caption{Rounds to convergence of the distributed strategies for $N_{\text{AP}}\in\{10,50,100\}$ ($\text{H}=1$ top, $\text{H}=6$ bottom).}
    \label{fig:rounds_results}
\end{figure}

\begin{figure}[!t]
    \includegraphics[width=\linewidth]{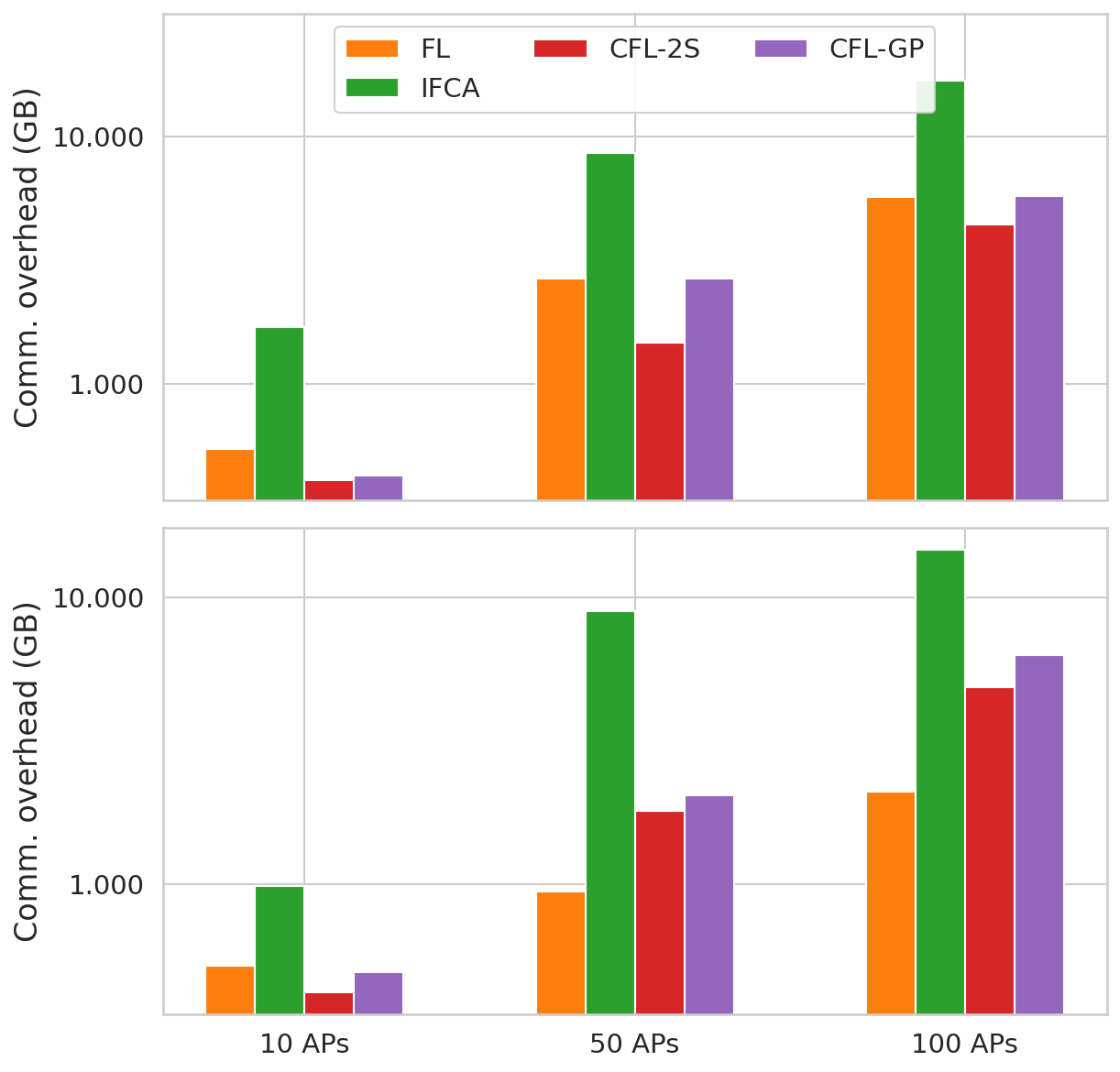}
    \vspace{-7mm}
    \caption{Communication overhead (log scale) of the distributed strategies for $N_{\text{AP}}\in\{10,50,100\}$ ($\text{H}=1$ top, $\text{H}=6$ bottom).}
    \label{fig:comm_results}
\end{figure}

\begin{figure}[!t]
    \includegraphics[width=\linewidth]{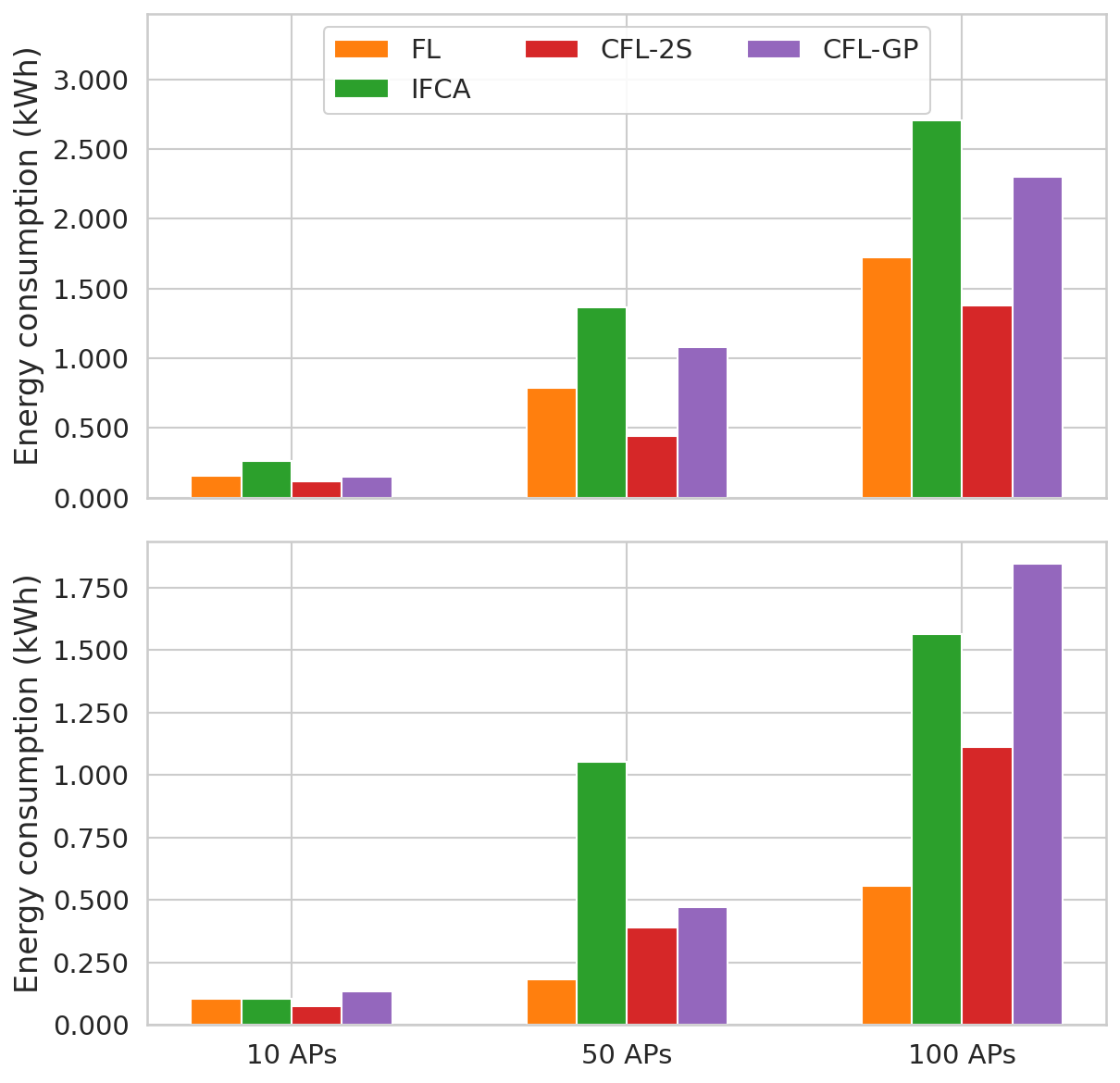}
    \vspace{-7mm}
    \caption{Computational energy of the distributed strategies for $N_{\text{AP}}\in\{10,50,100\}$ ($\text{H}=1$ top, $\text{H}=6$ bottom).}
    \label{fig:energy_results}
\end{figure}

Fig.~\ref{fig:rounds_results}, Fig.~\ref{fig:comm_results} and Fig.~\ref{fig:energy_results} report the convergence rounds, communication overhead, and energy-to-convergence, respectively, of the three \ac{cfl} strategies and \ac{fl} for both horizons and all network sizes, averaged over the 5 seeds. We examine each metric in turn, focusing on how \ac{cfl-2s} compares against the \ac{ifca}, \ac{cfl-gp} and \ac{fl} baselines.

\textbf{Rounds to convergence.} Convergence speed (Fig.~\ref{fig:rounds_results}) is the only dimension on which \ac{cfl-2s} is not consistently ahead among the clustered methods. At the short horizon, it reaches its best model in fewer rounds than \ac{ifca} at every network size and stays in a comparable range to \ac{cfl-gp}. 
At the longer horizon, instead, the ranking becomes mixed, with \ac{cfl-2s} occasionally needing more rounds, most notably more than \ac{cfl-gp} at the mid-sized network. 
Vanilla \ac{fl} sits at the two extremes: it is among the slowest at the short horizon (on par with \ac{ifca}), yet by far the fastest at the longer horizon for the larger networks, where its single global model plateaus early and triggers early stopping after fewer than half the rounds of any clustered method. As the next two metrics make clear, however, a higher round count does not penalises \ac{cfl-2s}.
Indeed, our proposal exchanges and processes far less information per round compared to the other \ac{cfl} methods, so more rounds do not translate into higher communication or energy.

\textbf{Communication overhead.}
Among the clustered methods, \ac{cfl-2s} achieves the lowest communication overhead in every configuration (Fig.~\ref{fig:comm_results}, logarithmic scale), exchanging on average ${\sim}73\%$ less data than \ac{ifca} and ${\sim}20\%$ less than \ac{cfl-gp}. The gap is by far the widest against \ac{ifca}, which must broadcast all $K$ cluster models to every \ac{ap} at each round so that each can select its own. The margin over \ac{cfl-gp} is narrower.
Indeed, both transmit only the relevant per-cluster model, but \ac{cfl-2s} adds no clustering-specific signalling, reusing the gradients already exchanged for training.
Crucially, these savings grow with the network size, precisely the regime of practical interest for managed Wi-Fi. 
Finally, vanilla \ac{fl} matches the per-round footprint of \ac{cfl-2s} but, converging over more rounds at the short horizon, transmits more in total. 
The sole exception is the longer horizon at the larger networks, where its early stopping makes it the cheapest of all.

\textbf{Computational energy.}
The same ordering carries over to energy among the clustered methods (Fig.~\ref{fig:energy_results}): \ac{cfl-2s} consumes the least energy-to-convergence in every configuration, on average ${\sim}49\%$ less than \ac{ifca} and ${\sim}37\%$ less than \ac{cfl-gp}. The reason is that its lightweight clustering and low per-round cost dominate the total budget, so a few extra rounds do not erode its lead.
At the longer horizon and the mid-sized network, for instance, \ac{cfl-2s} runs more rounds than \ac{cfl-gp} yet still spends the least energy. Which baseline is costliest is not fixed, however: \ac{ifca} dominates the energy budget at the short horizon, whereas \ac{cfl-gp} tends to be the most expensive at the longer one. Vanilla \ac{fl} mirrors its communication behaviour, falling between \ac{cfl-2s} and the other clustered methods at the short horizon but becoming the most energy-effienct option at the longer horizon for the larger networks.

Overall, among the clustered strategies \ac{cfl-2s} pairs the best \ac{mae} with the lowest communication and energy in every setting, making it a strong fit for resource-constrained managed Wi-Fi deployments. Relative to \ac{fl}, however, this advantage is conditional: \ac{cfl-2s} dominates at the short horizon and the smallest network, whereas \ac{fl}'s early-stopping global model is cheapest at the longer horizon for large networks---where \ac{cfl-2s} roughly doubles communication and energy to roughly halve the \ac{mae}. Clustering is therefore best invoked selectively, when \ac{ap} heterogeneity justifies its overhead.

\section{Conclusions}
\label{sec:conclusions}

This paper addressed traffic load prediction in centrally
managed Wi-Fi networks via a novel two-stage CFL methodol
ogy: Stage 1 generates and filters candidate AP partitions using
SVD-based gradient embeddings and a set of clustering quality
criteria; Stage 2 selects the partition maximizing the Gaussian
differential entropy of the smallest cluster, promoting well
trained models for the data-scarce groups most susceptible to
generalization degradation.

On a real-world campus Wi-Fi dataset, \ac{cfl-2s} achieves the best \ac{mae} among all distributed strategies, reducing it by up to $51\%$ over vanilla \ac{fl} and $21\%$ over \ac{cfl-gp}, with the largest gains at the longer horizon and larger networks. Still, in some cases, \ac{fl} is more energy-efficient and consumes less communication resources. This highlights that clustering pays should be invoked selectively, especially  when \ac{ap} data heterogeneity is high.

Future work will study sensitivity to hyperparameters and clustering quality criteria across more diverse tasks, and cost-aware selective clustering that activates dedicated per-cluster models only when their accuracy gain justifies the additional energy and/or communication cost.

\bibliographystyle{IEEEtran}
\bibliography{IEEEabrv,biblio}

\end{document}